\documentclass[12pt,a4paper]{article}

\usepackage{graphicx}
\usepackage{url}
\usepackage{amsmath}
\usepackage{setspace}
\usepackage{color}
\usepackage{float}
\usepackage{soul}

\usepackage{tikz}
\usetikzlibrary{shapes.geometric, arrows}
\tikzstyle{start} = [rectangle, rounded corners, minimum width=3cm, minimum height=1cm,text centered, draw=black, fill=red!30]

\tikzstyle{io} = [trapezium, trapezium left angle=70, trapezium right angle=110, minimum width=3cm, minimum height=1cm, text centered, draw=black, fill=blue!30]

\tikzstyle{process} = [rectangle, rounded corners, text centered, draw=black, fill=orange!30, inner sep=2mm]
\tikzstyle{process2} = [rectangle, rounded corners, text centered, draw=black, fill=blue!20, inner sep=2mm]
\tikzstyle{stop} = [rectangle, rounded corners, text centered, draw=black, fill=green!30, inner sep=2mm]

\tikzstyle{arrow} = [thick,->,>=stealth]

\usepackage{tabularx}

\graphicspath{{Images/}}

\begin{document}


\title{An empirical algorithm to forecast the evolution of the number of COVID-19 symptomatic patients after social distancing interventions}
\author{Luis Alvarez\medskip\\ 
         $^1$ CTIM. Departamento de Inform\'atica y Sistemas, \\
         Universidad de Las Palmas de Gran Canaria. Spain
        }
\date{}

\maketitle

\begin{abstract}
We present an empirical algorithm  to forecast the evolution of the number of COVID-19 symptomatic patients in the early stages of the pandemic spread and after strict social distancing interventions. The algorithm is based on a low dimensional model for the variation of the exponential growth rate that decreases
after the implementation of strict social distancing measures. From the observable data given by the number of tested positive, our model estimates the number of infected hindcast introducing in the model formulation the incubation time. We also use the model to follow the number of infected patients who later die using the registered number of deaths and the distribution time from infection to death.  The relationship of the proposed model with the SIR models is studied. Model parameters fitting is done by minimizing a quadratic error between the data and the model forecast. An extended model is also proposed that allows a longer term forecast. An online implementation of the model is avalaible at \url{www.ctim.es/covid19}
\end{abstract}


\section{Introduction}

In this work we propose an  empirical parametric model to forecast the evolution of the number of  
COVID-19 symptomatic patients, $N(t)$, after  social distancing interventions. 
This study presents  a numerical analysis of the effect of the confinement phase of the pandemic. It  attempts to predict the evolution of the  number of cases and deaths, based on past observations and assuming that the social distancing policy is steady or evolves slowly.  Hence the  main assumptions we  take are:

\begin{enumerate}
\item The evolution of the cumulative number of contaminated patients, $y(t)$, grows at an
exponential rate (that we name $a$), during a period of time $t_{0}$. We thus have $y'(t)=ay(t)$. Then,
after  social distancing measures are imposed the exponential rate $r(t)$ (such that $y'(t)=r(t)y(t)$) decreases until it attains
the value 0 at time $t_{1}.$ In this study, we considered first the following type of evolution for the exponential rate:
\begin{equation}
r_{1}(t)=\left\{
\begin{array}
[c]{ccc}%
a & if & 0\leq t\leq t_{0}\\
b & if & t\in(t_{0},t_{1}]\\
0 & if & t>t_{1}%
\end{array}
\right.  \label{eq:r(t)}%
\end{equation}%
but we realized that the next parametric model, with the same number of unknowns,  was more flexible and accurate:
\begin{equation}
r(t)=\left\{
\begin{array}
[c]{ccc}%
a & if & 0 \leq t \leq t_{0}\\
a\left(\frac{t_1-t}{t_{1}-t_{0}}\right)  ^{\gamma} & if & t\in
(t_{0},t_{1}]\\
0 & if & t>t_{1}.%
\end{array}
\right.  \label{eq:a(t)}%
\end{equation}
In the first model the parameters for $r_{1}(t)$ are $a,b,t_{0}$ and $t_{1}$, and the parameters
for $r(t)$ are $a,\gamma,t_{0}$ and $t_{1}$. The values for $a,b$ and $\gamma$ are always 
positive. The larger the value of  $\gamma$  the stronger  the effect of the  social distancing 
measures on the growth of $N(t)$.

\item The evolution of  the  number  $N(t)$ of the  symptomatic patients at  time $t$ depends  on  the evolution  law of  contaminated patients, and on the  law  of the  incubation  period.

\item At the beginning of the epidemic outbreak, the data of tested positive patients provided by most countries can be assumed to concern mostly symptomatic patients. This is a reasonable
assumption in the countries where tests were performed only on patients which show
some symptoms. It is important to point out that the available
databases about the coronavirus expansion make no distinction between infected
subjects which show symptoms or not. If we assume that the number of symptomatic 
patients is proportional to the number of registered infected subjects, the model still works. 
This is a reasonable assumption as long as a country keeps the same  infection test policy. 
If a country changes its testing policy and starts testing  more subjects, then many 
non-symptomatic subjects are going to be
included in the dataset, which can strongly deteriorate the accuracy of any observational model. This is why our  model  \eqref{eq:a(t)} for the decay of the exponential rate is  merely empirical, and aims at the  simplest formulation possible.

\item The  social distancing measures are taken at the beginning of the epidemic and there are many more exposed subjects  than infected and recovered
ones,  so that  we can  assume  that the variation of the  symptomatic patients only depends on the existing
contaminated  patients and the influence of the social distancing measures (see below the relation with the SIR model). 
\end{enumerate}

\bigskip

Regarding the distribution of the incubation period, Lauer et al. in
\cite{La20}, using the data of 181 patients approximate the distribution of
the incubation period as a log-normal distribution. The cumulative
distribution function of this log-normal is given by%
\begin{equation}
F(t)=\left\{
\begin{array}
[c]{ccc}%
\int_{0}^{t}\frac{e^{-\frac{(\log x-\mu)^{2}}{2\sigma^{2}}}}{x\sigma\sqrt
{2\pi}}dx & if & t>0\\
0 &  & \text{otherwise}%
\end{array}
\right.
\label{eq:F(t)}
\end{equation}
with $\mu=$ $1.621$ and $\sigma=0.418.$

The rest of the paper is organized in the following way: in section 2 we study the solution of equation (\ref{eq:y(t)}). In section 3, we analyze the relation of this model with the usual SIR model. In section 4, we present a short discussion about the lack of reliability of the available data of the COVID-19 spread. In section 5, we present the algorithm proposed to fit the model to the data. In section 6, we present an extension of the empirical  model to the forecast of the number of deaths. Section 7 presents the experimental setup. Finally  section 8 concludes.

\section{The empirical evolution model}

The continuous version of the evolution of contaminated subjects, $y(t)$, following an
exponential grow, $r(t)$, is given by the very basic differential equation: 

\[
y^{\prime}(t)=r(t)y(t).
\]
This equation can be solved explicitly, and in the case of $r(t)$ given by (\ref{eq:a(t)}) the solution is 

\begin{equation}
y(t)=Ce^{\int_{0}^{t}r(s)ds}=\left\{
\begin{array}
[c]{ccc}%
Ce^{at} & if & t\in\lbrack0,t_{0}]\\
Ce^{at_{0}}e^{\frac{a}{\gamma+1}\allowbreak\left(  (t_{1}-t_{0})-\left(
\frac{t_{1}-t}{t_{1}-t_{0}}\right)  ^{\gamma}\left(  t_{1}-t\right)  \right)
} & if & t\in\lbrack t_{0},t_{1}]\\
Ce^{at_{0}}e^{\frac{a}{\gamma+1}\allowbreak(t_{1}-t_{0})} & if & t>t_{1},
\end{array}
\right.
\label{eq:y(t)}
\end{equation}
the asymptotic state of the number of contaminated subjects is
\begin{equation}
Lim_{ t\rightarrow \infty} \  y(t)=Ce^{at_{0}}e^{\frac{a}{\gamma+1}(t_{1}-t_{0})},
\label{eq:asymp}
\end{equation}
and it is attained at $t_1$. Therefore the impact of the  social distancing measures is determined by the value : 
\begin{equation}
M_{a,\gamma,t_1,t_0}=\frac{a}{\gamma+1}(t_{1}-t_{0}).
\label{eq:M}
\end{equation}
The smaller this value, the more effective the social distancing interventions. We notice that the peak in the new daily contaminated patients is obtained when $y'(t)$ changes sign which corresponds to $y''(t)=0$.  Using a straightforward computation we obtain that the peak is attained at 

$$
t_{peak}=t_{1}-\left(  \frac{\gamma}{a}(t_{1}-t_{0})^{\gamma}\right)  ^{\frac{1}{\gamma+1}}
$$
The evolution of symptomatic subjects, taking into account the cumulative distribution of the incubation time, $F(t)$, is given by 
\begin{equation}
N(t)=\int_{0}^{t}y^{\prime}(s)F(t-s)ds.
\label{eq:N(t)}
\end{equation}
We observe that since $F(t)$ converges to 1 when $t$ goes to $\infty$, then 

\[
Lim_{t\rightarrow\infty}N(t)=Ce^{at_{0}}e^{\frac{a}{\gamma+1}\allowbreak
(t_{1}-t_{0})}-C.
\]%
Notice that $y(0)=C$, $N(0)=0$ and there is a delay between the evolutions of $y(t)-C$ and $N(t)$, corresponding to the time required by contaminated subjects to become symptomatic. Notice that the official number of cases is related to $N(t)$  which is the variable that can be observable. $N(t)$ does not follow, in general an exponential growth. In fact: 
\[
a\approx\log
\frac{y(t+1)}{y(t)}\neq \log\frac{N(t+1)}{N(t)}=\log\frac{\int_{t_{0}}^{t_{1}}y^{\prime
}(s)F(t+1-s)ds}{\int_{t_{0}}^{t_{1}}y^{\prime}(s)F(t-s)ds}.
\]%

In practice, the observable data is the number of registered tested positive patients. An extra time is required from the moment the patient shows symptoms until the test is done and it is finally recorded as tested positive. This time strongly depends on the in-country logistics. In this work we assume that this time is about 2 days, so in expression \ref{eq:N(t)} we replace $F(t-s)$ by $F(t-s-2)$. This modification does not change the profile of  $N(t)$, it  simply represents an extra delay between the evolution of contaminated subjects and the evolution of the registered tested positive subjects. 

\subsection{An extended model to track different trend modifications}

The exponential growth given by equation (\ref{eq:a(t)}) is very simple and it is useful to compute an estimation of $y(t)$ after an strict lockdown is implemented, this estimation covers from the epidemic outbreak until a certain time after the daily peak. However if we want to go further and to approximate the evolution for a longer time we need to extend the model to have more flexibility in order to fit the epidemic spread. In fact, the above basic model can be easily extended in the following way:  let $\{t_0^k,t_1^k\}_{k=1,..,K}$ be 2 increasing sequences of real numbers satisfying $t_0^{k+1}<t_1^k$. $\{t_0^k\}$ represent times where a change is expected in the evolution trend of the epidemic. Then the exponential growth (\ref{eq:a(t)}) can be extended in the following way: 


\begin{equation}
r(t)=\left\{
\begin{array}
[c]{ccc}%
a_1 & if & t\in\lbrack0,t_{0}^1]\\
a_k\left(  \frac{t_{1}^k-t}{t_{1}^k-t_{0}^k}\right)  ^{\gamma_k} & if & t\in(t_{0}^k
,t_{0}^{k+1}] \quad k=1,2,..,K-1\\
a_K\left(  \frac{t_{1}^{K}-t}{t_{1}^{K}-t_{0}^{K}%
}\right)  ^{\gamma_{K}} & if & t\in(t_{0}^{K},t_{1}^{K}]\\
0 & if & t>t_{1}^{K}.%
\end{array}
\right.  \label{eq:r(t)_extended}%
\end{equation}
Notice that $r(t)$ can be discontinuous at $t_{0}^{k}$ because a  relaxation of social distancing  measures  will definitely produce an abrupt modification in the exponential growth of the epidemic. We point out that the function $r(t)$ is always decreasing except at the possible points of discontinuity $t_{0}^{k}$ . In particular, the model is not well adapted to scenarios where the growth rate can grow continuously, such as a second epidemic wave.

\section{Relation with the SIR model}
The basic SIR model separates the population in three compartments: $S(t)$ (the
number of susceptible), $I(t)$ (the number of infectious), and $R(t)$ (the
number of recovered). It should be  mentioned in this model that the number of dead is  negligible.  We  can also consider that R(t) is the sum of recovered and deceased. Each member of the population typically progresses from
susceptible to infectious to recovered. The basic SIR model to estimate
$S(t),$ $I(t)$ and $R(t)$ is the following system of ordinary differential
equations:
\[%
\begin{array}
[c]{l}%
\frac{dS}{dt}=-\beta\frac{I}{I+S+R}S\\
\frac{dI}{dt}=\left(  \beta\frac{S}{I+S+R}-\gamma\right)  I=\left(  R_0\frac{S}{I+S+R}-1\right) \gamma I\\
\frac{dR}{dt}=\gamma I,
\end{array}
\]
where $\beta$ and $\gamma$ are parameters which depend on the particular
disease. $R_0=\frac{\beta}{\gamma}$, named  the reproductive number, is one of the key parameters in transmission models and it represents the number of secondary infections that arise from a typical primary case in a completely susceptible population.  Notice that $S(t)$, the number of  susceptible subjects, is a decreasing function. When the ratio between $S(t)$ and the total population satisfies 
$$
\frac{S(t)}{I(t)+S(t)+R(t)}=\frac{1}{R_0},
$$
we obtain  $\frac{dI}{dt}(t)=0$. Hence the peak of infected subjects is attained, and from that time, the number of infected subjects  starts decreasing. Notice that the larger $R_0$, the larger the time required to attain the infection peak. We observe that in our model, in the evolution of contaminated patients, $y(t)$, we include the infected and recovered subjects, so $y(t)=I(t)+R(t)$ and then using the SIR model we obtain that

\begin{equation}
\frac{dy}{dt}(t)=  \beta \frac{S(t)}{I(t)+S(t)+R(t)}(y(t)-R(t)).\\
\label{eq:SIR_y(t)}
\end{equation} 
The SIR model with constant $\gamma$ and $\beta$  and the conclusions about the peak of infected subjects make sense only if the virus propagates freely across  time,  but everything changes if we impose  social distancing measures to the population. A natural way to include human interventions in the SIR model is to replace $\beta$ by a time dependent  function $\beta(t)$. This strategy has been used by different authors in different contexts using extended versions of the SIR models. For instance in \cite{Ch04}, the authors propose the following exponential type function: 
\[
\beta(t)=\beta_{1}+(\beta_{0}-\beta_{1})e^{-q(t-t_{0})_{+}}%
\]
another exponential type function has been introduced in \cite{LIU2021110501}:
$$
\tau(t)=\tau_{1}e^{-\mu(t-N)_{+}}
$$
In \cite{PiZa20} the following rational function is proposed:  
\[
\beta(t)=\beta_{0}(1-\rho(t-t_{0})/t)
\]
In \cite{Ga20} the author proposes the function
\begin{equation}
\tau(t)=\tau_{0}[1-\mu(t-N)_{+}]_{+}.
\label{eq:linear}%
\end{equation}
We observe that this is a particular case of the function \eqref{eq:a(t)} defining $r(t)$ 
where $a=\tau_{0},$ $\mu=1/(t_{1}-t_{0})$, $N=t_{0}$ and $\gamma=1$. The only difference of this function with $r(t)$ is that in $r(t)$ we add the power $\gamma$ to modulate the way the exponential growth rate decreases.

In this work we assume that social distancing interventions govern  the evolution of contaminated subjects rather than 
the SIR dynamic and we replace equation (\ref{eq:SIR_y(t)}) by 

\[
\frac{dy}{dt}(t)=r(t)y(t).
\]
Therefore we include in the term $r(t)$ the impact of the human interventions, the influence of the ratio between $S(t)$ and the total population and the influence of $R(t)$. The latter makes sense if we are at the beginning of the pandemic (so $R(t) \approx 0$)  or if we assume that $R(t)$ is proportional to $y(t)$. By focusing just on the number of contaminated subjects we reduce the complexity of the problem and we avoid to deal with the balance between infected, exposed and recovered patients which is very difficult to estimate properly due to the lack of accuracy in the data we can manage about the number of infected subjects. We point out that in our model $y(t)$ is the number of infected patients which show symptoms, which is the data most countries provide when using PCR tests.


\section{A discussion about the reliability of the existing data about the coronavirus expansion in terms of the evaluation of the impact of  social distancing interventions}

{\bf Tested  positive subjects:} First, we stress again that what we can observe is the evolution of tested positive subjects, which is quite different from the evolution of contaminated subjects. This value strongly depends on the testing policy which can change across the time. If the testing policy does not change too much during the period  of time used to estimate the model, our  forecast will still  be valid to some  extent. This value has the advantage that it is the first one to react to the installation of  social distancing measures. \medskip

{\bf \noindent  Symptomatic tested  positive subjects:}  With the existing variety of testing policies, this value seems to be more reliable than just tested positive subjects. On the one hand official data make no distinction between symptomatic and asymptomatic tested positive subjects. On the other hand, when the health system is overwhelmed, many symptomatic subjects not requiring  hospitalization are  simply sent home without  testing.  \medskip 

{\bf \noindent  Number of deaths:} Theoretically, this is a reliable data, but when the health system is overwhelmed a significant number of patients  die without being counted as affected by the coronavirus, so the accuracy of this data depends on the capacity of the health system to properly count  the deaths. This is far from being the case when the health system is completely overwhelmed. \medskip 

{\bf \noindent Number of hospitalizations or number of patients in intensive care:} Again, theoretically, these data are more reliable than the number of tested positive, but again, in the case of a health system completely overwhelmed, the quality of these data is strongly deteriorated. Another issue with these data is the way they are provided. In some cases, the official data refer to the current situation where the patients which leave the hospital or reanimation are removed from the statistics. \medskip 

Another important issue in the data quality is the time required for a new case to be included in official statistics. For example, if new PCR positive tests and new antibody positive tests are added at the same day, the quality of the data deteriorates seriously. Indeed,  both detection correspond to infections at very different past times!  Even using only PCR tests, the time from the presentation of symptoms to inclusion in official statistics must be taken into account. In Spain, this time is distributed with a median of 6 days;   in 25\% of cases it is even more than 10 days. This delay deteriorates the usability of the data, and hinders a short-term prediction of the evolution of the epidemic.

\section{The algorithm }
As discussed in the previous section, the data we use are far for being reliable. In our approach we use a very simple model with few  parameters in the hope that the simplicity of the model can compensate in some way for the lack of accuracy of the data and provide a big picture of the evolution of pandemic expansion after social distancing interventions. We observe that if we have not enough data after the implementation of social distancing measures, the parameters $\gamma$ and $t_1$ cannot be computed properly from the data. To reduce the uncertainty in the calculation of these parameters, we can set "a priori" the expected value of the effectiveness of the containment  effectiveness given by $M_{a,\gamma,t_1,t_0}$, defined in (\ref{eq:M}), based on the values obtained for other countries which have implemented previously the same kind of social distancing  measures. That is, we can constrain the effectiveness to satisfy 

\begin{equation}
M_{a,\gamma,t_1,t_0}=\frac{a}{\gamma+1}(t_{1}-t_{0})=M_0
\label{eq:M2}
\end{equation}
where $M_0>0$. 

\subsection{Model discretization}

We estimate $N(t)$ by discretizing equation (\ref{eq:N(t)}) in the following basic way

\begin{equation}
N(t) = \int_{0}^{t}y^{\prime}(s)F(t-s)ds \approx \sum_{k=0}^{k=t_1-1} (y(k+1)-y(k))F(t-(k+0.5)),
\label{eq:N(t)D}
\end{equation}
where $y(t)$ is given by  (\ref{eq:y(t)}).

\subsection{Parameter adjustment}
Given a dataset, $D(t)$, of the number of symptomatic patients across the time
for a region, we fix the parameters by minimizing the quadratic mean error%
\begin{equation}
Error_{[t_{min},t_{max}]}(C,a,\gamma,t_{0},t_{1},\tilde{t})=\frac{1}{\sum_{t=t_{\min}}^{t_{\max}}w(t)}%
\sum_{t=t_{\min}}^{t_{\max}}w(t)\left(  D(t)-N(t+\tilde{t})\right)  ^{2},
\label{eq:error}
\end{equation}
where $\tilde{t}$ is the translation of $N(t)$ to fit $D(t).$ The interval $[t_{\min
},t_{\max}]$ is the range of values we use to fit the parameters of the model. We assign the  following weight $w(t)$ to  each data value in the model estimation: 

\begin{equation}
w(t)=(t-t_{\min}+1)^\alpha
\label{eq:w}
\end{equation}
where $\alpha \geq 0$. When $\alpha=0$ all points in the dataset have the same weight ($w(t)\equiv 1$).  
The higher the value of $\alpha$  the more weight it will be giving to the latest values of the dataset. To adjust the model parameters, we use a Newton-Raphson type method combined with an extensive search exploration of potential parameter interval values.

\medskip
\noindent {\bf Computation of C:} we observe that $C$ is a scale factor and if the other parameters of the model are given, $C$ can be estimated by equating to zero the derivative of the error (\ref{eq:error}) with respect to $C$, which yields the following expression for $C$: 

\begin{equation}
C=\frac{\sum_{t=t_{\min}}^{t_{\max}}w(t)D(t)N_{1}(t+\tilde{t})}{\sum_{t=t_{\min}}^{t_{\max}%
}w(t)N_{1}(t+\tilde{t})N_{1}(t+\tilde{t})}
\label{eq:C}
\end{equation}
where $N_{1}(t+\tilde{t})=N(t+\tilde{t})$ is computed using $C=1$ and the other given parameters. 
We point out that the values of $C$ and $t_0$ are very related, in terms of the evolution of contaminated subjects, $y(t)$. Indeed the values $C'=Ce^a$ and $t'_0=t_0-1$ provide the same results as $C$ and $t_0$. 

\subsection{Computing the minimum of $Error_{[t_{min},t_{max}]}(C,a,\gamma,t_{0},t_{1},\tilde{t})$}

First, we observe that, in general, due to the strong variation of the
available data, the quadratic error, $Error_{[t_{min},t_{max}]}(C,a,\gamma,t_{0}%
,t_{1},\tilde{t})$ can have several local minima. To avoid getting trapped in
a spurious local minimum, we use a basic optimization strategy, where we
combine massive evaluations of $Error_{[t_{min},t_{max}]}(C,a,\gamma,t_{0},t_{1},\tilde{t})$ in 
large discrete intervals with a basic Newton-Raphson type method  to 
improve locally $a,\gamma,\tilde{t}$. To simplify the complexity, we use (\ref{eq:C}) to
express $C$ as a function of the rest of parameters, that is, $C\equiv
C(a,\gamma,t_{0},t_{1},\tilde{t}),$ so the quadratic error becomes
$Error_{[t_{min},t_{max}]}(a,\gamma,t_{0},t_{1},\tilde{t})\equiv Error_{[t_{min},t_{max}]}(C(a,\gamma,t_{0}%
,t_{1},\tilde{t}),a,\gamma,t_{0},t_{1},\tilde{t})$. The times $t_0$ and $t_1$ are computed in integer precision and the rest of parameters in floating precision. The computation of $t_0$ and $t_1$ in integer precision has little influence on the final result because small variations of $t_0$ are mostly compensated modifying $C$ and small variations of $t_1$ are mostly compensated modifying $\gamma$.  

\bigskip

\paragraph{\bf MAIN STEPS OF THE OPTIMIZATION  ALGORITHM}

\begin{itemize}
\item Step 1: Computation of $t_{\min}$ and $t_{max}$. We define $t_{\min}$ as the the time
when the data starts to grow with an exponential growth with a minimum value of
10, that is: 
\[
t_{\min}=\min\{t:D(t-2)>10\text{ and }D(t)>1.1D(t-1)\text{ and }%
D(t-1)>1.1D(t-2)\}
\]
and we fix $t_{\max}=min\{t_{min}+N_1,t_c\}$, where $N_1$ is a parameter of the algorithm to fix the number of days used to compute the model. $t_c$ is the max available time in the data set observation.

\item Step 2: Initial estimation of $\tilde{t}.$ We fix initially the
following reference values for the rest of parameters: $a=0.13,$ $\gamma=2,$
$t_{0}=17$ and $t_{1}=52$. Then $\tilde{t}$ is computed initially in integer
precision as
\[
\tilde{t}_{0}=\underset{k\in N\cap\lbrack k_{\min},k_{\max}]}{\arg\min
}Error_{[t_{min},t_{max}]}(a,\gamma,t_{0},t_{1},k)
\]
where $[k_{\min},k_{\max}]$ has been fixed experimentally as $[k_{\min
},k_{\max}]$ =$[-26-t_{\min},10-t_{\min}].$

\item Step 3: Computing an initial minimum evaluating the energy in parameter
intervals. For each parameter $p\in\{a,\gamma,t_{0},t_{1},\tilde{t}\}$ we
define a discrete interval $I_{p}=\{p_{1},..,p_{N_{p}}\}$ (in the case of
$p=\tilde{t}$, $I_{p}$ is a neighborhood of $\tilde{t}_{0}$ computed above) and
we define the set $\mathcal{I}=I_{a}\times I_{\gamma}\times I_{t_{0}}\times
I_{t_{1}}\times I_{\tilde{t}}.$ We compute a first minimum, $P_{0},$ as
\[
P_{0}=\underset{(a,\gamma,t_{0},t_{1},\tilde{t})\in\mathcal{I}}{\arg\min
}Error_{[t_{min},t_{max}]}(a,\gamma,t_{0},t_{1},\tilde{t})
\]
This "brute force" technique has the advantage that it can be easily implemented using parallelization and to a certain extent, it avoids getting trapped in spurious local
minima. Once $P_{0}$ is computed, it is improved using a basic Newton-Raphson method to optimize $a,\gamma,\tilde{t}$.

\item Step 4. Improving iteratively the minimum location: For $k=0,1,2,..$ we
use a small discrete neighborhood, $N_{P_{k}},$ of $P_{k},$ and we define $P_{k+1}$ as 
\[
P_{k+1}=\underset{(a,\gamma,t_{0},t_{1},\tilde{t})\in N_{P_{k}}}{\arg\min
}Error_{[t_{min},t_{max}]}(a,\gamma,t_{0},t_{1},\tilde{t})
\]
after this initial estimation, $P_{k+1}$ is improved using the Newton-Raphson method. 
Iterations stop when
\[
\frac{Error(P_{k})-Error(P_{k+1})}{Error(P_{k})}<TOL
\]
where $TOL$ is a convergence parameter (we fix $TOL=10^{-6}$ in the algorithm
implementation). This iterative procedure allows to improve the minimum
estimation. In particular, it allows the minimum to go beyond the initial
parameter interval $\mathcal{I}$.


\end{itemize}

As quoted before, at the beginning of the epidemic spread, when not much data is available it can be useful to fix the expected value of effectiveness of the containment  effectiveness given by $M_{a,\gamma,t_1,t_0}$ defined in (\ref{eq:M}), in that case the value of $M_{a,\gamma,t_1,t_0}$ becomes a parameter of the algorithm and this value constraints the parameter optimization steps of the algorithm.

\subsection{Adaptation of the algorithm to the extended model}
In the case where the exponential growth is given by the extended model (\ref{eq:r(t)_extended}),  we compute the unknowns of the model, given by $C$, $\tilde{t}$ and $t_0^k,t_1^k,a_k,\gamma_k$ for k=1,..,K, in the following way: 

\begin{enumerate}
    \item We compute $C$, $\tilde{t}$ and $t_0^1,t_1^1,a_1,\gamma_1$ using the algorithm explained above.
    \item for each k=2,..,K, we compute $t_0^k,t_1^k,a_k,\gamma_k$ iteratively in the following way: 
    \begin{itemize}
        \item  $t_{max}=min\{t_{min}+\sum_{i=1}^{k}N_i,t_c\}$. (where $N_k$ is a parameter of the algorithm to fix how many days we consider to compute the model $k$)
         \item $t_0^k=\tilde{t}+t_{max}-N_k$
         \item We compute $t_1^k,a_k,\gamma_k$ by minimizing the quadratic error in the interval $[t_{min},t_{max}]$ with respect to $t_1^k,a_k$ and $\gamma_k$. 
    \end{itemize}
    
\end{enumerate}

\section{Forecasting the number of deaths}

We can easily extend the model to the case of the evolution of the number of deaths. In this case $y(t)$ represents the number of contaminated subjects who die and $N(t)$ the registered number of deaths. The only thing we have to change is the cumulative distribution $F(t)$. In that case we have to use the  infection-to-death time distribution. In \cite{FlIC20}, the authors model this distribution as the sum of two independent random times, both being  Gamma distributed with mean 5.1 days and coefficient of variation 0.86 and 18.8 days and a coefficient of variation 0.45 respectively.  The infection-to-death distribution is therefore given by

$$
i_{frm} \cdot (Gamma(5.1,0.86) + Gamma(18.8,0.45))
$$
where $i_{frm} $ is the population averaged over the age structure of a given country. As $i_{frm}$  is a constant factor, we can assume that $i_{frm}=1$ because in our model, this factor will be compensated by the constant factor $C$. Therefore, changing, in expression (\ref{eq:N(t)}), $F(t)$ by the cumulative distribution of the  infection-to-death time distribution we can follow the evolution of the number of deaths.

\begin{figure}[ptb]
\begin{center}
\includegraphics[width=0.75\linewidth]{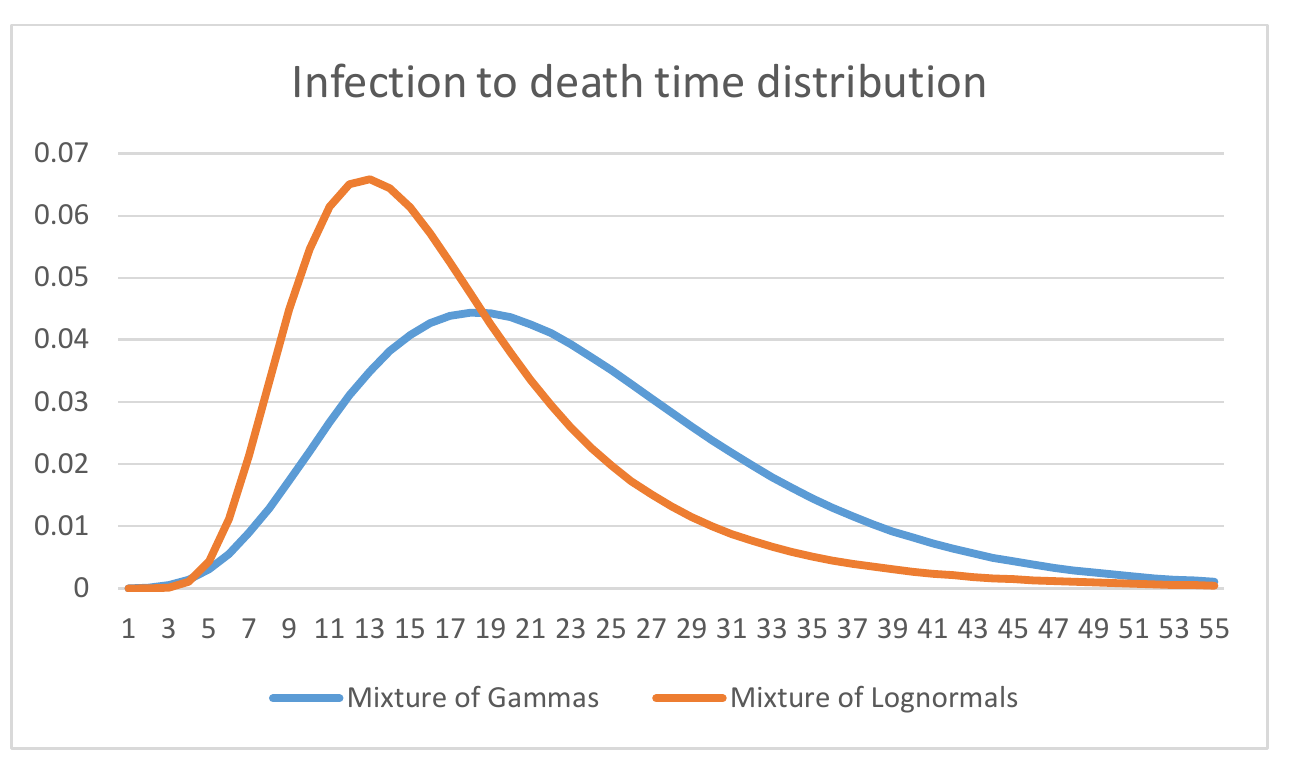}
\end{center}
\caption{ Comparison of the infection to death time distribution using the mixture of Gammas proposed in \cite{FlIC20} and the one obtained using the mixture of lognormals (\ref{eq:lognormal}). }%
\label{fig:IntectedToDeathDistribution}%
\end{figure}

\begin{figure}[ptb]
\begin{center}
\includegraphics[width=1\linewidth]{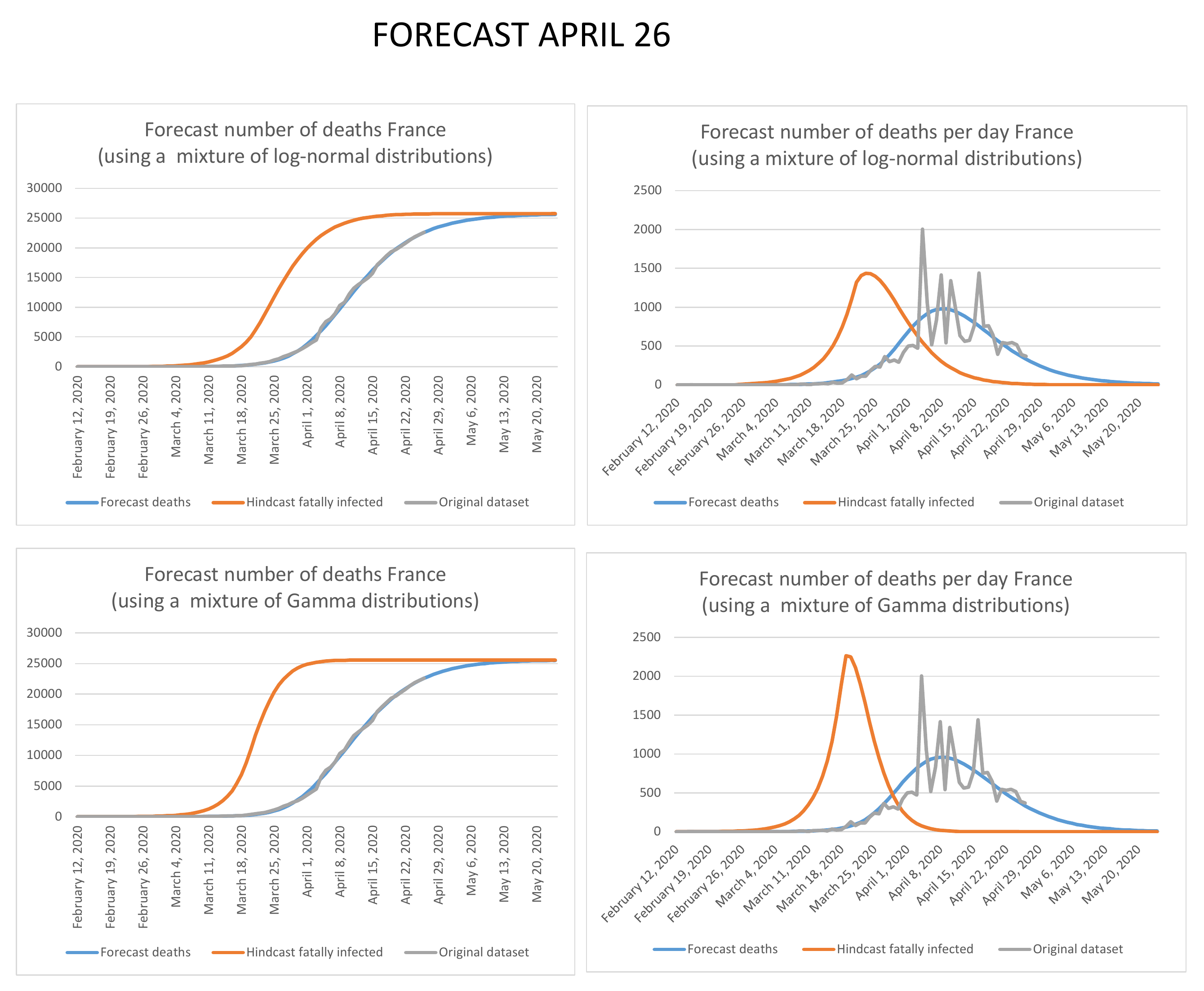}
\end{center}
\caption{ Comparison of the forecasts obtained by the proposed model using 
the infection to death time distribution proposed in \cite{FlIC20} and the one obtained using (\ref{eq:lognormal}). }%
\label{fig:ForecastFranceDeathApril25}%
\end{figure}

In the same way, assuming that we know the time distribution of other registered values as for instance,  the infection-to-hospitalization time distribution, we can forecast, using the same model, the evolution of the corresponding  registered value. We point out that the time distribution of the COVID-19 registered values is a topic under investigation and the results can change in the next future. For instance, the study presented in  \cite{Sa20} suggests that there are two sub-populations in delays between hospitalization and death: individuals that die quickly upon hospital admission (15\% of fatal cases, mean time to death of 0.67 days) and individuals who die after longer time periods (85\% of fatal cases, mean time to death of 13.2 days). The combination of Gamma distributions presented above does not reflect this behavior.     In the official   Spain report
\cite{ReSp20}  using the information of 9765 patients, it is estimated that, in the case of men, the time from the onset of symptoms to death has a median of $11$ days with quartiles $Q1=7$ and $Q3=16$. In the case of women, these values are median$=10$, $Q1=6$ and $Q3=14$. Based on the values for men an women, we approximate, experimentally,  the distribution of the time from the onset of symptoms to death as a $lognormal(\mu=2.351375257,\sigma=0.6011434688)$ distribution. The median of this distribution is $=10.5$, $Q1=7$ and $Q3=15.75$. So we can approximate the distribution of the time from infection to death as the following mixtures of lognormal distributions 
\begin{equation}
\text{lognormal}(\mu=1.621,\sigma=0.418) \ + \ \text{lognormal}(\mu=2.351375257,\sigma=0.6011434688),
\label{eq:lognormal}
\end{equation}
where the first one corresponds to the time infection to the onset of symptoms (see (\ref{eq:F(t)})). In Fig. \ref{fig:IntectedToDeathDistribution}, we compare the profile of the distributions using the mixture of Gammas proposed in \cite{FlIC20} and the one obtained using the mixture of lognormals (\ref{eq:lognormal}). We point out that they are quite different.    Using the mixture of Gammas, a patient takes
considerably more time to die from the infection. In Fig. \ref{fig:ForecastFranceDeathApril25} we compare the forecasts obtained by the proposed model using 
the infection to death time distribution proposed in \cite{FlIC20} and the one obtained using (\ref{eq:lognormal}). We observe that the forecast of deaths are quite similar but the forecasts of fatally affected subjects are very different. We believe that the one obtained by the lognormals is more plausible because in the other one the number of fatally affected subjects goes to zero too quickly with respect to the evolution of deaths. In the IPOL online demo we use the one obtained by the mixture of lognormals. However, we believe that this approximation is not very accurate  either, and  as quoted before, we think that the knowledge and accuracy of  the time distribution of the basic epidemic factors will be improved in the near future.

\newpage

\section{Experimental setup}

We invite the reader to use the online interface available in \url{www.ctim.es/covid19} to get an idea of the performance of the model.  The user can apply it to study the evolution of the first wave of the epidemic in any country in the world. We point out that the model is designed to adjust to an exponential evolution of the number of infected where the growth rate is always decreasing (except in some potential isolated points of discontinuity generated when the extended model is applied). Therefore, the model is well adjusted to approximate the evolution of a first wave but it is not suitable for modeling the change in trend of a second wave where the exponential growth rate starts to rise again.

\subsection{Description of the online interface parameters}
\label{subsection:parameters}

The parameters we use in the  online DEMO interface are the following: 

\begin{enumerate}
    \item {\em Type of data}: it can be tested positive or deaths.
    \item {\em Number of days to compute the basic model}: this parameter is denoted by $N_1$ in the text. It represents the number of days used to compute the model after the number of cases starts to grow exponentially (that is $t_{min}$). 
    \item {\em Constraining lockdown effectiveness}: if this option is activated by the user, then she/he can constrain the lockdown effectiveness given by equation (\ref{eq:M}). 
    \item {\em Use extended model to fit trend modification}: if this option is activated by the user, then she/he can choose the number of days used to compute two extra extended models. These parameters are denoted by $N_2$ and $N_3$ in the text. If the value of one of these parameters is zero, then no extended model is computed. Using two extended models and the basic model (the first one) we can manage situations where the pandemic outbreak initially starts to growth exponentially and then, due to lockdown measures the growth rate starts to decrease (this is managed by the basic model), then the growth rate changes its trend, because, for instance, the test capacity of the country improves (this can be managed by the first extended model) and finally the evolution stabilizes around a baseline (this can be managed by the second extended model). Many  countries have followed these  three phases when a strict lockdown has been implemented. We point out that the model we propose is not expected to simulate properly the impact of mild social distancing measures or of a second wave.
    \item {\em Weight in least squares fitting}: the parameter $\alpha$ in equation (\ref{eq:w}). 
    \item {\em The country or uploaded data used} . 
\end{enumerate}

\section{Conclusions}

The proposed algorithm for the basic model is able to forecast quite well the evolution of the epidemic spread in its early stage when little information is available and strict social distancing measures are implemented. If we fix manually the value of $M_{a,\gamma,t_{0},t_{1}}$ using the one obtained for other countries where similar social distancing measures have been implemented we can improve the results in a significant way. We have experimentally estimated, using the cases of Italy, Spain and France,  that the value of $M_{a,\gamma,t_{0},t_{1}}$ when a strict lockdown is applied is bewteen $1.4$ and $1.7$. 

The formulation of the extended model allows us to properly track the evolution of the full course of the first epidemic wave.  

A critical  reader will have noticed that  contrarily to the SIR  models,  our model based on  the  $r(t)$ law  is empirical. This is justified by three facts that  we  have stressed: 

a) Given the huge observation noise it is better  to work with  a  very  low dimensional model, so that we estimate a  very  few empirical parameters,  rather than the  many that cannot actually be estimated; 

b) As highlight in the paper, to cope with the influence of the social distance measures, SIR models also require to define empirical models for the growth rate $\beta(t)$. 

c) The  virtue of the proposed empirical model is  that  it  may cope not  only with noise,  but also with a variation  of the  very  definition of observed variables.  This variation definitely happens. Indeed, the  various  administrations are progressively changing  the way they make their statistics about the observed cases.  They also adapt their  testing  policy, and ultimately  they also adjust their containment policy.   Thus,  an adapted parametric approach to the prediction might be adequate to overcome all these limitations.

\clearpage

\bibliographystyle{siam}

\end{document}